\documentclass[]{article}
\usepackage{amsmath,bbold,graphicx,array,slashed,amssymb,multirow,cite}
\usepackage[vcentermath]{youngtab}
\usepackage{pstricks}
\usepackage{color}

\newcommand{\Tr}{\text{Tr}}

\title{Supersymmetric and non-supersymmetric Seiberg-like dualities for gauged Wess-Zumino-Witten theories, realised on branes}
\author{E. Ireson\\
~\\
{\it Department of Physics, Swansea University}\\
{\it Singleton Park, Swansea, SA2 8PP, UK}\\
{\texttt{pyireson@swansea.ac.uk}}}

\begin{document}

\maketitle

\begin{abstract}
In this work we extend the results of previous derivations of Seiberg-like dualities (level-rank duality) between gauged Wess-Zumino-Witten theories. The arguments in use to identify a potential dual for the supersymmetric WZW theory based on the coset $\frac{U(N+M)_k}{U(N)_k}$ can be extended to be applied to a wider variety of gauge groups, notably $\frac{USp(2N+2M)_{2k}}{USp(2N)_{2k}}$ and $\frac{SO(2N+2M)_{2k}}{SO(2N)_{2k}}$, which will be dealt with briefly. Most interestingly, non-supersymmetric versions of the latter theories can also be shown to have duals in a similar fashion. These results are supported by several pieces of evidence, string phenomenological interpretations of Seiberg duality, even in non-supersymmetric backgrounds, is helpful to justify the formulation, then, from field theory, quantities such as central charges or Witten indices are shown to match exactly. The stability of these non-supersymmetric models is also discussed and shown to be consistent.
\end{abstract}
\clearpage
\section{Introduction}

The use of branes and assemblages thereof has proven to be a particularly fruitful area of investigation in which to probe properties of string theory, field theory, and the bridges that exist between them. In particular, the Hanany-Witten construction \cite{Hanany:1996ie} provides an ideal laboratory to embed field theories in string theory, the properties of which we can use as tools to propose, and to some extent justify, advanced concepts and notions in field theory. However, these models have the specificity of being supersymmetric by nature, and it is usually unclear if we can make analogous statements in non-supersymmetric scenarios. Progress has been made in that direction however: starting from some of the very simplest, explicitly non-supersymmetric set-ups (the Sugimoto model) \cite{Sugimoto:1999tx}, it has been argued that phenomena such as dynamical symmetry breaking (see \cite{Uranga:1999ib}, also\cite{Sugimoto:2012rt}) can still be made sense of, using analogous notions, which started interest in this direction. Meanwhile, a rich theory of brane constructions was developed, capable of reproducing several classes of field theories \cite{Giveon:1998sr}, encompassing their features and enabling us to formulate more of their properties, further examples include Chern-Simons theories \cite{Kitao:1998mf}, or theories with non-unitary gauge groups with the use of orientifolds \cite{Polchinski:1995mt}. There is much hope, then, that from all these models will eventually arise deeper notions that will further our understanding of particle theories.

It is our purpose here to continue the ongoing interest in non-supersymmetric generalisations of known results in these brane constructions. Much success has been had modelling Seiberg duality (and all of its variants e.g. Giveon-Kutasov duality \cite{Giveon:2008zn} etc.) in this matter, and it seems possible so far to perform such generalisations in a selected number of cases \cite{Armoni:2008gg},\cite{Armoni:2013ika},\cite{Armoni:2014cia}. Here we present more extensions of recently published results \cite{Armoni:2015jsa} which describe a variant of such Seiberg dualities in the case of gauged Wess-Zumino-Witten theories, constructed as defects in the brane picture. Using many of the very same tools used generically to analyse these situations, we will show that there are seemingly no obstacles in performing the generalisations in question.

\subsection{Recapitulation of previous results}

In a recent paper \cite{Armoni:2015jsa}, it was argued that specific gauged Wess-Zumino-Witten theories were dual to each other, in the sense of Seiberg. These theories are built from the following basic action: let $M$ be a three dimensional space with a boundary $\Sigma$, let $G,H\subset G$ be Lie groups and $A,B$ be gauge fields in the Lie algebras thereof, $\omega_3$ the Chern-Simons three-form on $M$ ($\omega_3(A)=AdA+\frac{2}{3}A^3$), finally let $g$ be a field valued in $G$ (it is a group element, not a Lie Algebra element) on $\Sigma$ and $\Pi_{\mathcal{H}}$ the projection on the Lie Algebra of $H$:

\begin{align}\label{WZWaction}
S=\frac{k}{4\pi}\int_M \left( \omega_3(A^g)-\omega_3(\Pi_\mathcal{H}B^g) \right) -&\frac{k}{8\pi}\int_\Sigma \left( \Tr(g^{-1}D_Ag)^2-\Tr\left(\Pi_\mathcal{H}(g^{-1}D_Bg) \right) \right),\\
\text{ where }A^g&=g^{-1}Ag-g^{-1}dg\nonumber
\end{align}

We have somewhat abused notation in the above: $g$ is technically only a boundary degree of freedom (i.e. defined on $\Sigma=\partial M$). However, it can be extended uniquely to a field over the whole of $M$: we will here be dealing with groups that satisfy $\pi_2(G)=0$ \cite{Witten:1983ar} such that this extension is possible. On the other hand, since generically for these groups $\pi_3(G)=\mathbb{Z}$, the action itself (namely, the first part, over $M$) is only well-defined mod $2\pi$, leading to the quantisation conditions on $k$, which will henceforth be an integer.

This action may seem a little protracted and obscure, in particular it is not obvious that it is a boundary action, but we can write it more explicitly. We move to light-cone coordinates, which makes the $A_+,B_+$ become Lagrange multipliers. Eliminating them imposes $F_{2-}=G_{2-}=0$ (these quantities being the field strengths for $A,B$). This can be solved for by choosing $A_i=U^{-1}\partial_i U \text{ , }B_i=V^{-1}\partial_i V\text{ , }i=-,2$ for $U,V$ Lie Algebra elements of $G$ and $H$. We then define new variables $h,\tilde{h}$ and $a_{\pm}$ such that
\begin{equation}
\Pi_{\mathcal{H}}(Vg)=h^{-1}\tilde{h}\text{ , }Ug=h^{-1}g\tilde{h}, a_- =\partial_- \tilde{h} \tilde{h}^{-1}\text{ , }a_+=\partial_+ h h^{-1}
\end{equation}

With these conventions, and judicious use of the Polyakov-Wiegmann Identity \cite{Polyakov:1984et}, it can be shown \cite{Armoni:2015jsa} that the total action can be written as:

\begin{align}
S&=-\frac{k}{8\pi}\int_{x^2=0} d^2x \Tr\left(g^{-1}\partial g\right)^2 -\frac{k}{12\pi}\int_{x^2\geq 0}d^3x \Tr\left(g^{-1}\partial g\right)^3 \nonumber\\
&+ \frac{k}{4\pi}\int_{x^2=0}\Tr\left( -a_+g^{-1}a_-g+a_+g^{-1}\partial_- g -\partial_+ g g^{-1}a_-+a_+a_-\right) 
\end{align}

This is already somewhat clearer: the first two terms are readily recognisable as a standard WZW action \cite{Witten:1983ar}, the extra term, on the larger space, is purely topological, non-propagating. It serves a specific purpose though, it allows for a very particular form of symmetry for the theory.

Indeed, this action possesses an unusual symmetry, namely a partially gauged group. WZW models generically have $G_L\times G_R$ left- and right-action global symmetry, but the construction at hand involves an enhanced $H$ local symmetry of the left-action, leaving only the coset of transformations $\frac{G}{H}$ as global. To denote such theories, we will simply just refer to the coset. These types of theories can be made supersymmetric, even off-shell, without spoiling this gauge invariance \cite{Armoni:2015jsa}. As detailed in \cite{Witten:1983ar}, this effectively accounts for the presence of the boundary space without needing to impose any specific boundary conditions, we have simply added extra fields to account for the effects thereof.

Then, the precise statement of \cite{Armoni:2015jsa} was that the $\mathcal{N}=(1,0),(2,0),(1,1)$ supersymmetric theories of coset

\begin{equation}
\frac{U(N+M)_k}{U(N)_k}
\end{equation}

had  "Seiberg-like" (Giveon-Kutasov-like) duals in similarly supersymmetric theories with coset

\begin{equation}
\frac{U(k-N)_k}{U(k-N-M)_k}
\end{equation}

It is worth reminding the reader that purely bosonic WZW (and CS) theories obey a level-rank type duality, which can be proven very thoroughly (matching the conformal primaries, for instance, in WZW theories \cite{DiFrancesco:1997nk}), in which the dual rank and level are the level and rank (respectively) of the original theory. The proposition above is manifestly different, this is due to the presence of UV fermion degree of freedom. In the far IR, removing these fermionic degrees of freedom produces one-loop shifts in the level of the theory (further loop effects are forbidden by the integer nature of $k$), so that the final, purely bosonic dual pair are indeed a level-rank pair. This effect is not dependent on supersymmetry, only on the appearance of fermions in any representation, which will still lead to consistent results, see \cite{Armoni:2014cia}. 

This duality was discussed in several ways in the previous work \cite{Armoni:2015jsa}, the most elaborate of which was translating the problem into string theory constructions. Seiberg duality has a very concrete interpretation in string theory as a brane creation/annihilation phenomenon as a result of swapping specific fivebranes \cite{Hanany:1996ie}, this effect was argued to orchestrate the duality in this case also, like it is in a number of cases. Furthermore, field-theoretical checks of the duality such as the computation of the Witten Index and the matching of the central charges of the theories were put forward to justify the duality. These clues are quite non-trivial in their matching, giving good evidence to support the construction process.  

Now, these constructions open up the way to proving further instances of the same phenomenon, which we will detail here: firstly, through the use of orientifold objects in string theory, the case of supersymmetric theories based on the orthogonal ($SO$) and symplectic ($USp$) subgroups of $U(N)$ can be dealt with along the same lines, but secondly and most surprisingly that these subgroup theories can be made non-supersymmetric in specific ways without spoiling the duality altogether. This is done in the same spirit as already established non-supersymmetric dualities, see \cite{Sugimoto:2012rt},\cite{Armoni:2013ika},\cite{Armoni:2014cia}, that is, using a collection of anti-branes and an orientifold to break all of supersymmetry, such that the resulting field theory is still stable and does not contain e.g. tachyons. Then, tools previously used to analyse the purported dualities can be used to justify once again the proposition at hand, even in a non-supersymmetric case (specifically, matching of the central charge in the theory).

\subsection{Further dualities}\label{s1}

We would like to argue, by the same arguments already in use in an established case, that the following two-dimensional theories form a dual pair:

\begin{table}[!h]
\centering
\begin{tabular}{ccc}
 A gauged Wess-Zumino model of the coset
  & &
A gauged Wess-Zumino model of the coset 
 
\\
\Large{$\frac{USp(2N+2M)_{2k}}{USp(2N)_{2k}}$ }&\LARGE{$\simeq$} & \Large{$\frac{USp(2k-2N-2)_{2k}}{USp(2k-2N-2M-2)_{2k}}$}\\

with $\mathcal{N}=(1,0),(2,0),(1,1)$ supersymmetry & & with $\mathcal{N}=(1,0),(2,0),(1,1)$ supersymmetry \\

\multicolumn{3}{c}{\phantom{and}} \\

 A gauged Wess-Zumino model of the coset
  & &
A gauged Wess-Zumino model of the coset 
 
\\
\Large{$\frac{SO(2N+2M)_{2k}}{SO(2N)_{2k}}$} &\LARGE{$\simeq$} & \Large{$\frac{SO(2k-2N+2)_{2k}}{SO(2k-2N-2M+2)_{2k}}$}\\

with $\mathcal{N}=(1,0),(2,0),(1,1)$ supersymmetry & & with $\mathcal{N}=(1,0),(2,0),(1,1)$ supersymmetry \\
\end{tabular}
\end{table}

But also that the following explicitly non-supersymmetric theories are dual:

\begin{table}[!h]
\label{nsd1}
\centering
\begin{tabular}{ccc}
 A gauged Wess-Zumino model of the coset
  & &
A gauged Wess-Zumino model of the coset 
 
\\
\Large{$\frac{USp(2N+2M)_{2k}}{USp(2N)_{2k}}$} &\LARGE{$\simeq$}& \Large{$\frac{USp(2k-2N+2)_{2k}}{USp(2k-2N-2M+2)_{2k}}$}\\

with antisymmetric fermionic currents && with antisymmetric fermionic curents \\

\multicolumn{3}{c}{\phantom{and}} \\
 A gauged Wess-Zumino model of the coset
  & &
A gauged Wess-Zumino model of the coset 
 
\\
\Large{$\frac{SO(2N+2M)_{2k}}{SO(2N)_{2k}}$} &\LARGE{$\simeq$}& \Large{$\frac{SO(2k-2N-2)_{2k}}{SO(2k-2N-2M-2)_{2k}}$}\\

with symmetric fermionic currents && with symmetric fermionic currents \\
\end{tabular}
\caption{Stating the new purported dualities}
\label{props}
\end{table}

And finally the following non-supersymmetric Type-0'B theories form dual pairs:
\clearpage
\begin{table}[!h]
\centering
\begin{tabular}{ccc}
 A gauged Wess-Zumino model of the coset
  && 
A gauged Wess-Zumino model of the coset 
 
\\
\Large{$\frac{U(2N+2M)_{2k}}{U(2N)_{2k}}$} &\LARGE{$\simeq$}& \Large{$\frac{U(2k-2N+2)_{2k}}{U(2k-2N-2M+2)_{2k}}$}\\

with antisymmetric fermionic currents && with antisymmetric fermionic currents \\

\multicolumn{3}{c}{\phantom{and}} \\
 A gauged Wess-Zumino model of the coset
  & &
A gauged Wess-Zumino model of the coset 
 
\\
\Large{$\frac{U(2N+2M)_{2k}}{U(2N)_{2k}}$} &\LARGE{$\simeq$}& \Large{$\frac{U(2k-2N-2)_{2k}}{U(2k-2N-2M-2)_{2k}}$}\\

with symmetric fermionic currents && with symmetric fermionic currents \\
\end{tabular}
\end{table}

The precise notion of the representation of bosonic and fermionic currents will be made clearer in section \ref{checks}, but briefly: though the 2D group-valued field $g$ (and its superpartner or would-be superpartner) is in a $G_L\times G_R$ bifundamental representation, the (anti-)holomorphic currents it generates are in a symmetrised two-index representation of the gauge group by construction.

Evidence for this proposed duality will be extracted from brane constructions but also verified directly via the computation of central charges for these CFTs and observing that they coincide, and, when possible, by direct computation of the Witten index. It is not yet known if any other relevant physical quantities of these dual pairs match, as so far only these tools were in use for the simplest case, the supersymmetric unitary case.

\section{Brane construction of boundary theories}

\subsection{Brane content}
We will construct the advertised theories using Type IIB brane assemblages involving $D3$ branes, $D$, $NS$ and composite fivebranes, and importantly an orientifold plane.

\begin{table}[h]
\centering
\begin{tabular}{|rr|cccccccccc|}
\hline
		 $NS5$& : & $0$ & $1$ & $2$   & $3$ & $4$ & $5$ &   &  &  &  \\ 
 $2N2+M$ $D3_+$ & : & $0$ & $1$ & $2_+$ &     &     &     & $|6|$ &  &  &  \\ 
    $2N$ $D3_-$& : & $0$ & $1$ & $2_-$ &     &     &     & $|6|$ &  &  & \\ 
 $O3$         & : & $0$ & $1$ & $2$ &     &     &     & $6$ &  &  &  \\
    $2N+2M$ $D5$& : & $0$ & $1$ &       & $3$ & $4$ & $5$ &  &  &   &\\ 
 $(1,2k)5$ (A) & : & $0$ & $1$ & $2$   & $3$ &     &$[\scriptsize{\begin{tabular}{c}5\\9\end{tabular}}]_{-\theta}$     &  &  & $8$  &  \\ 
\hline
\end{tabular}
\caption{Brane content of the $\mathcal{N}=(1,0)$ 2D theory}
\end{table}

In the above notation $|6|$ means finite extent in the 6 direction (the threebranes are bounded by fivebranes and stop at the junction), and $2_{\pm}$ means that the brane extends only along the positive (resp. negative) $2$ direction, stopping at the fivebranes placed at $x_2=0$. 

The $(1,2k) 5$ brane is a composite object, bearing both $NS$ and $D$ charge. Created by the admixture of $1$ $NS$ and $2k$ $D5$ branes, it naturally stretches at an angle between the directions not shared between the $D$ and $NS$ branes, whose magnitude depends on the proportion of either brane type, here it is $\tan(\theta)=2k$. It imposes mixed boundary conditions on the stacks of threebranes connected to it, a linear combination of the $D$ and $NS$ boundary conditions, whose relative coefficients depend on $k$, such that the extremal cases $k=0$ and $k\rightarrow \infty$ subsume to $NS$ or $D$ respectively.  These new boundary conditions spoil the invariance of the action under variation of the gauge field, unless a specific boundary term  is added to compensate them, as has been shown in \cite{Kitao:1998mf}. This term leads to the appearance of a Chern-Simons piece in the low-energy action. For all intents and purposes, it possesses the important features of its component pieces, this can be justified further using a construction of the composite brane in M-theory as was done in the paper cited above.

Finally, the $Op$ orientifold is not strictly speaking a dynamical object, but the invariant subspace of the following world-sheet and space-time projection in Type IIB string theory. Let $z=\exp(\tau+i \sigma)$,

\begin{equation}
\Omega:x^M(z,\bar{z})\rightarrow -x^M(\bar{z},z)\text{ , }M=p+1,...9
\end{equation}

We take the string theory resulting in imposing definite symmetry under this inversion. This object has a variety of properties (explained in more detail in e.g.  \cite{Johnson:2000ch}): it un-orients the strings, and identifies the string modes above and below the invariant plane. The plane itself is just a geometrical object, it does not have dynamics of its own, and is usually pencilled into depictions of brane constructions as a visual reminder of this action on the string modes, nevertheless it can be shown that for most intents and purposes it can be thought of as preserving the same set of Killing spinors as the equivalent $Dp$ brane (given its effect on the space-time coordinates) and effectively having $Dp$ brane charge, so that it is indeed useful to include it in the details of the construction like any other brane.

 When we draw this invariant plane to lie along a stack of $N$ branes, upon which we look at the field-theory modes, it projects the full $U(N)$ gauge symmetry of the stack of branes down to one of its two subgroups of definite symmetry under transposition: $SO(2N)$ or $USp(2N)$, which possess antisymmetric and symmetric (symplectic) generators respectively. These groups are materialised on the brane stack by using an extra set of $N$ mirror branes at the orientifold. However, motion of these mirror branes are forced to match the motion of the original branes by the presence of the plane: whilst there are effectively $2N$ units of brane charge, they can only move in a symmetric fashion away from the plane, reducing the number of degrees of freedom one would naively expect from such a setup. Effectively, only $N$ of these branes are dynamical, which is natural since there were $N$ dynamical branes in the original model. This is valid for all types of $D$ branes in the theory, in particular it applies to the $D5$ branes participating in the creation of the $(1,2k)$, explaining why we impose that this level is strictly even. We will use the convention of labelling it $2k$ whenever orientifolds are present, but in its absence it could be any integer.

The content above has been shown to lead to an $\mathcal{N}=(1,0)$ 2D theory, but with a few modifications of the extent of the fivebranes we can also generate an $\mathcal{N}=(1,1)$ or $\mathcal{N}=(2,0)$, see Table \ref{branes}.

\begin{table}[!h]
\centering
\begin{tabular}{|rr|cccccccccc|}
\hline
		 $NS5$& : & $0$ & $1$ & $2$   & $3$ & $4$ & $5$ &   &  &  &  \\ 
 $2N+2M$ $D3_+$ & : & $0$ & $1$ & $2_+$ &     &     &     & $|6|$ &  &  &  \\ 
    $2N$ $D3_-$& : & $0$ & $1$ & $2_-$ &     &     &     & $|6|$ &  &  & \\ 
 $O3$         & : & $0$ & $1$ & $2$ &     &     &     & $6$ &  &  &  \\
    $2N+2M$ $D5$& : & $0$ & $1$ &       &  & &  & $6$ & $7$  & $8$   & $9$\\ 
 $(1,2k)5$ (A) & : & $0$ & $1$ & $2$   & $[\scriptsize{\begin{tabular}{c}3\\7\end{tabular}}]_{-\theta}$  &     &    &  &  & $8$  & $9$  \\ 
\hline
		 $NS5$& : & $0$ & $1$ & $2$   & $3$ & $4$ & $5$ &   &  &  &  \\ 
		 $2N+2M$ $D3_+$ & : & $0$ & $1$ & $2_+$ &     &     &     & $|6|$ &  &  &  \\ 
		 $2N$ $D3_-$& : & $0$ & $1$ & $2_-$ &     &     &     & $|6|$ &  &  & \\ 
		 $O3$         & : & $0$ & $1$ & $2$ &     &     &     & $6$ &  &  &  \\
		 $2N+2M$ $D5$& : & $0$ & $1$ &       & $3$ & $4$ & $5$ & $6$  &  &   &\\ 
		 $(1,2k)5$ (A) & : & $0$ & $1$ & $2$   & $[\scriptsize{\begin{tabular}{c}3\\7\end{tabular}}]_{-\theta}$  &     &    &  &  & $8$  & $9$  \\
		 \hline 
\end{tabular}
\caption{Brane content of the $\mathcal{N}=(1,1)$ and $\mathcal{N}=(2,0)$ 2D theories respectively}
\label{branes}
\end{table}
\clearpage
Constructions of these types have been previously argued \cite{Armoni:2015jsa} to lead to gauged WZW models on the space of the junction between the two threebrane stacks. Indeed, it is known that threebranes connecting an $NS$ and a $(1,k)$ fivebrane have (at sufficiently low energies) a Chern-Simons theory $U(N)_k$ living on their worldvolume \cite{Kitao:1998mf}. The junction acts as an explicit boundary for physical space, which in turn generates WZW degrees of freedom in the junction. To see this, we can look at open string modes between the different branes, and explain what happens when branes reconnect.

\begin{figure}[h]
	\centering
			\includegraphics[height=0.32\textheight]{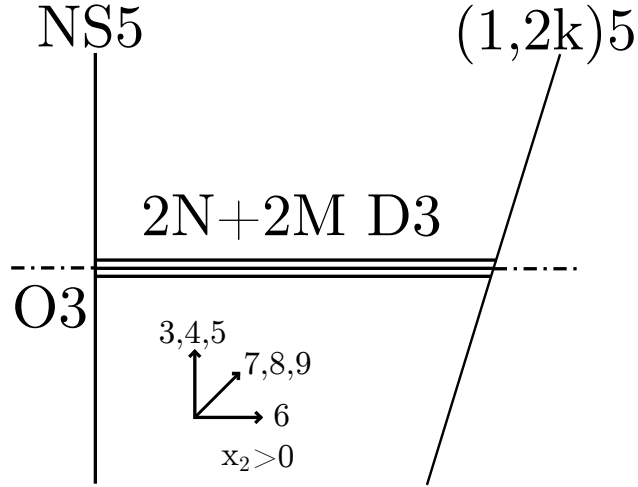}
	\caption{Part of construction at fixed $x_2$ (setup identical at $x_2<0$ but for having $2N$ $D3$ instead)} 
\end{figure}

\begin{figure}[h]
	\centering
			\includegraphics[height=0.33\textheight]{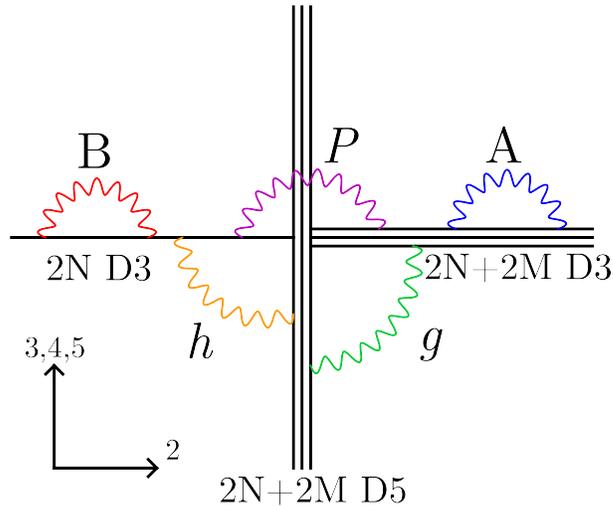}
	\caption{The junction at $x_2=0$, and open string modes generating the required boundary theory} \label{fig:modes}
\end{figure}

\clearpage
In detail, the string modes are represented in Figure \ref{fig:modes}. The total symmetry group of this boundary construction is $G_1 \times G_2 \times H$, the factors of which are, in order, the symmetry group of the larger $D3$ stack, the symmetry group of the $D5$ stack, and that of the smaller $D3$ stack. Because the $D5$ and larger $D3$ stack both have $2N+2M$ branes, it happens that $G_1=G_2$  ($=G$, aligning with our previous notation), but their group actions on the various fields are different, as is the action of $H$ which is in effect always a subgroup of $G$. In detail here is a list of all the relevant modes and their representation under the aforementioned groups, we have:

\begin{itemize}
\item A gauge field $A$ in the adjoint of $G_1$,
\item A gauge field $B$ in the adjoint of $H$,
\item A bifundamental $g$ of $G_1\times G_2$,
\item A bifundamental $h$ of $H\times G_2$,
\item A bifundamental $P$ of $H\times G_1$.
\end{itemize}

Generically, the two brane stacks are separated in space, ending at different points of the boundary. The inter-stack distance corresponds geometrically to a mass for the $P$ modes: in this circumstance, $P$ is massive and decouples, effectively disconnecting the gauge groups of the two stacks.

However, we wish to align the stacks and reconnect them through the boundary. This happens naturally in the $\mathcal{N}=(1,0)$ theories since this direction is not a modulus, but the higher supesymmetric theories need to be fixed to this sector. We impose that the stacks align, in this configuration, $P$ is now massless and its lowest (scalar) component gets a vacuum expectation value of the form
\begin{equation}
\left<P|_{\theta=0}\right>=\left(\begin{array}{cccc}
1 &  &  &  \\ 
 & ... &  & 0 \\ 
 &  & 1 & 
\end{array} \right)
\end{equation}

 which breaks the $H\times G_1$ gauge symmetry to the diagonal $H_{\mbox{diag}}$, enforcing communication between the sectors. 

Furthermore, there exists an extra interaction at the boundary: \cite{Armoni:2015jsa}

\begin{equation}
\mathcal{W}=\Tr \left( g^\dagger P^\dagger h + h^\dagger P g\right) 
\end{equation}

which in general induces quartic interactions between the scalars in the multiplet. This potential clearly has a flat direction whenever $g=h$. If the scalar in $P$ gets a diagonal vacuum expectation value in its first $N$ components, the superpotential above will generate quadratic terms of interaction i.e. mass terms for $g$ and $h$, for all field configurations other than the ones at the flat direction. This therefore imposes that 

\begin{equation}
h=\Pi_{\mathcal{H}}(g)
\end{equation}

Thus we recover precisely the modes in the action we stated at the very beginning (Eq. \ref{WZWaction}). This argument should, without loss of generality, be valid for a unitary theory (for which it was originally formulated) or with an orientifold: being non-dynamical there are no new fields or interactions that are created due to its presence, it only serves to reduce the symmetry group.

\subsection{Seiberg duality and the absence of supersymmetry}
Seiberg duality (and all others related to it, as is the case here) has a natural interpretation in terms of fivebranes crossing other fivebranes. As has been explained in other works (see \cite{Giveon:1998sr} for a comprehensive review), this particular brane operation is expected to leave certain linear combinations of total brane charges invariant under the swap. This charge conservation imposes that the number of threebranes connecting the swapped fivebranes changes in a predictable fashion under this swap: after the swap, for every brane joining the $NS$ and the $D$ fivebranes, an antibrane is created, which annihilates the brane, effectively disconnecting the outer fivebranes; for every unconnected $D5$, a brane is created to connect it to the $NS5$. An analogous statement can be made for stacks of antibranes. Again one merely needs to consider the complicated $(1,k)5$ as a combination of $k$ $D5$ and $1$ $NS5$ in this instance, and it has been argued that the statement still holds (giving rise to Giveon-Kutasov duality \cite{Giveon:2008zn}).  

One needs an extra step when dealing with orientifolds: indeed, an $O3$ has the same charge (up to a sign) as two $D3$ branes \cite{Johnson:2000ch}, this charge generates additional brane creation/annihilation during the swap. A positively charged orientifold, like the one we have here, will count like 2 extra branes existing in the electric theory, thus leading to more antibranes being created, or equivalently more branes being annihilated.  

The final step into producing a construction that possesses the advertised field theories (namely, the non-supersymmetric versions of the Type IIB constructions found in section \ref{nsd1}) is to turn the $D$ branes into \textit{anti-branes}. Because the orientifold preserves the same supersymmetry as a normal brane, this eliminates all supercharges. From the point of view of field theory, this is manifested by a difference in representation between the fermionic and bosonic degrees of freedom. The orientifold action acts differently on them, if the bosonic modes are (for instance) symmetrised, the fermionic modes are projected onto their antisymmetric components \cite{Schwarz:2001sf}. This mismatch in representation between the components of the would-be multiplets is an explicit supersymmetry breaking. In this case, one may wonder if the string picture has any validity whatsoever: non-supersymmetric string theories may generically possess tachyons, or very undesirable loop effects. It has been argued \cite{Sugimoto:2012rt},\cite{Armoni:2013ika} that this particular method of supersymmetry breaking does lead to tractable field theories, at least for positively charged orientifolds, as they are expected to attract anti-branes. Negatively charged orientifolds would repel the anti-branes, and so we will proceed to explain in section \ref{stab} why our propositions are still valid.
\subsection{Type 0'B theory}

The third type of dual pairs discussed in this paper come from Type 0'B string theory. They are constructed as a particular orientifold of Type 0B theory (see \cite{Sagnotti:1995ga} \cite{Sagnotti:1996qj},\cite{Armoni:2013kpa}). Type 0 string theory is non-supersymmetric by nature but possesses a tachyon, thankfully the orientifold procedure eliminates it. The resulting field theories are characterised by unitary gauge groups, but with Dirac gauginos (which can be split up into two Majorana) in a representation with definite symmetry, this is to be expected given that supersymmetry should be restored in the large N limit, in which only planar diagrams survive.

\clearpage
\subsection{Reading off the results from the brane construction}

Finally, we synthesize the previous statements to explain the results stated in Section \ref{s1}. They fit in three broad classes of models, as detailed at the beginning in Table \ref{props}.

\begin{itemize}
\item Supersymmetric models with orientifolds. As explained, the effect of the orientifold is to reduce the symmetry groups of the branes, allowing us to make statements about $SO$ and $USp$ groups, but we have to take into account the fact they have brane charge, thus affecting the dual group rank. $O3^+$ leading to $USp$ gauge groups, and counts as extra branes, hence a shift of $-2$ appears. The opposite happens for $O3^-$.
\item Non-SUSY models with orientifolds. Now, $O3^+$ annihilates additional branes under the swap, which translates to the creation of anti-branes, hence the shift is $+2$ this time. This shift portrays the non-supersymmetric nature of the duality in field theory.
\item Type-0'B models. The non-supersymmetric nature comes from the construction of the string background, which is all characterised using Sagnotti's particular orientifold \cite{Sagnotti:1995ga}\cite{Sagnotti:1996qj}. No anti-branes are needed in this case, so $O3^-$ begets a positive shift and conversely. 
\end{itemize}

We will now detail how and why these effects are perceived in the field theories borne on the threebranes of the construction, thus justifying to some extent the statements read from the brane picture.

\section{Field theoretical checks of the purported dualities}\label{checks}

A most valuable tool to check this duality is, as in \cite{Armoni:2015jsa}, to verify that the central charges match on either side. Particularly it is important to underline the non-triviality of the result: were it a particular feature of the general forms of the central charges in question, it would be less notable. As it stands there is very little reason to expect that these theories have matching central charges, since they involve groups of very different ranks. We will also mention that the supersymmetric theories have matching Witten indices (which count the number of supersymmetric vacua of the theory). 

\subsection{The WZW-Matter central charge formula}

As in the original paper \cite{Armoni:2015jsa}, the central charges (or gravitational anomalies) of the electric and magnetic theories can be computed and found to be equal, but one must ensure that they are computed taking general representations for the bosonic and fermionic fields.

The particle content of the 2D theory is that of a bosonic WZW model in an adjoint representation $R_B$, along with charged free fermions in a potentially different representation $R_F$, thereby generating holomorphic currents, $J(z)$ and $j(z)$ respectively.

First let us remind ourselves of the Sugawara construction \cite{Sugawara:1967rw} for the purely bosonic WZW theory. Let $k$ be the (tree-level) level of the theory, $f^{abc}$ are the structure constants of the symmetry group at hand:
\begin{equation}
J^a(z)J^b(w)\approx\frac{k\delta^{ab}}{(z-w)^2}+i\frac{f^{abc}J_c(w)}{z-w} \\
\end{equation}

Thereby, the OPE for the energy-momentum tensor in each sector can be computed (see \cite{DiFrancesco:1997nk} for a detailed computation, suppressed here), up to an important normalisation factor.

\begin{equation}
T_B(z)=\gamma J^a(z)J^a(z)\text{ , }\gamma=\frac{1}{k+C_2(R_B)} 
\end{equation}

Note the appearance of the factor of $C_2(R_B)$ (the dual Coxeter number for the representation of the bosons): it betrays quantum effects in the theory. Indeed the action we consider is obtained as the infra-red of a higher theory, where the gauge degrees of freedom ($A$,$B$) are allowed to propagate (via a Yang-Mills term). Being massive, as we go towards low energies, they will generically cease to propagate, but the effective theory at this scale must still bear a trace of what once was, that is, ultraviolet shifts in the level.

With these expressions, the bosonic central charge of the WZW model is then

\begin{align}
c_B(R_B)&=\frac{k}{k+C_2(R_B)}\dim (R_B)=\left(1-\frac{C_2(R_B)}{k+C_2(R_B)} \right)\dim (R_B) 
\end{align}

This is the result of the Sugawara construction, a well-known expression found for instance in \cite{DiFrancesco:1997nk}

Now let us add the fermionic currents in the picture. The fermionic superpartners to the bosons in this theory are actually free, so they have the following simple OPE and energy-momentum:
\begin{align}
j^a(z)j^b(w)&\approx\frac{k\delta^{ab}}{(z-w)}\\
T_F(z)&=\gamma^\prime j^a(z)j^a(z)\text{ , }\gamma^\prime =\frac{1}{k}
\end{align}

Leading to the following central charge:
\begin{align}
c_F(R_F)&=\frac{1}{2}\dim(R_F)
\end{align}

However, adding supersymmetry does not simply result in superposing the two sectors. Now, in addition to the gauge degrees of freedom, their superpartners also lead here to a further shift of the level, negative this time by virtue of its fermionic origin, for much the same reason as previously.

\begin{equation}
k\rightarrow k-C_2(R_F)
\end{equation}

Which altogether leads to the following corrected central charges in a WZW-matter system:

\begin{align}
c_B(R_B)&=\left(1-\frac{C_2(R_B)}{k+C_2(R_B)-C_2(R_F)} \right)\dim (R_B) \\
c_F(R_F)&=\frac{1}{2}\dim(R_F)
\end{align}

In a supersymmetric case, $R_B=R_F(=R)$, the corrections cancel and therefore we arrive at the following total central charge

\begin{equation}\label{susyc}
c(R)=\left(\frac{3}{2}-\frac{C_2(R)}{k}\right)\dim(R)
\end{equation}

Which is the usual result, found e.g. in \cite{Gaiotto:2013gwa}.

Finally, to move to a gauged WZW system based on the coset $\frac{G}{H}$ the total central charge is simply

\begin{equation}
c=c_G-c_H
\end{equation}

Where $c_G,c_H$ are the total (fermionic+bosonic) central charges explained above.

With these formulae, it is possible to observe that all of the proposed dualities have consistent central charges, which is a non-trivial process, since they involve different gauge group ranks on either side. Let us briefly summarise the results.

\subsection{Supersymmetric theories}
We now present key figures in the identification of the central charges relevant for the supersymmetric dualities, showing the form of the various gauge-theoretical quantities needed to construct the central charge, in Tables \ref{tab:spsym} and \ref{tab:sosym}.

\begin{table}[h]
\centering
\begin{tabular}{|c|c|c|c|c|}
\hline Groups & {$USp(2N+2M)_{2k}$} & {$USp(2N)_{2k}$}& $USp(2k-2N-2)_{2k}$ &   $USp(2k-2N-2M-2)_{2k}$ \\ 
\hline {\Large$\phantom{\frac{U}{V}}$} $\dim(R)$ & $\frac{(2N+2M)(2N+2M+1)}{2}$  & $\frac{(2N)(2N+1)}{2}$  & $\frac{(2k-2N)(2k-2N+1)}{2}$  & $\frac{(2k-2N-2M-2)(2k-2N-2M-1)}{2}$ \\ 
\hline {\Large$\phantom{\frac{U}{V}}$} $C_2(R)$ & $2n+2M+2$ & $2N+2$ &$2k-2N$ & $2k-2N-2M$ \\ 
\hline \Large{$c_{\frac{G}{H}}$} &  \multicolumn{4}{c|}{\phantom{\LARGE{$\frac{U}{V}$}}$\frac{-M}{2k}(-3 k (1 + 2 M + 4 N) + 2 (1 + 2 M^2 + 6 N + 6 N^2 + M (3 + 6 N)))$}   \\ 
\hline 
\end{tabular}
\caption{Gauge-theoretic data for the central charge of the first supersymmetric duality}
\label{tab:spsym}
\end{table}

\begin{table}[h]
\centering
\begin{tabular}{|c|c|c|c|c|}
\hline Groups & $SO(2N+2M)_{2k}$ & $SO(2N)_{2k}$ & $SO(2k-2N+2)_{2k}$ & $SO(2k-2N-2M+2)_{2k}$  \\ 
\hline {\Large$\phantom{\frac{U}{V}}$} $\dim(R)$ & {$\frac{(2N+2M)(2N+2M-1)}{2}$}  & $\frac{(2N)(2N-1)}{2}$  & $\frac{(2k-2N)(2k-2N-1)}{2}$  & $\frac{(2k-2N-2M+2)(2k-2N-2M+1)}{2}$ \\
\hline {\Large$\phantom{\frac{U}{V}}$} $C_2(R)$ & $2N+2M-2$ & $2N-2$ &$2k-2N$ & $2k-2N-2M$ \\ 
\hline \Large{$c_{\frac{G}{H}}$} &  \multicolumn{4}{c|}{\phantom{\LARGE{$\frac{U}{V}$}}$\frac{M}{2k}(3 k (-1 + 2 M + 4 N) - 2 (1 + 2 M^2 - 6 N + 6 N^2 + M (-3 + 6 N)))$}   \\ 
\hline 
\end{tabular}
\caption{Gauge-theoretic data for the central charge of the second supersymmetric duality}
\label{tab:sosym}
\end{table}

The expression for the central charge is long and involves many terms, it is naively surprising that they match. For instance, from Table \ref{tab:spsym}, we insert the data into Eq. (\ref{susyc}) to find
\begin{align}
&\frac{1}{2} (2 M+2 N) (2 M+2 N+1) \left(\frac{3}{2}-\frac{2 M+2 N+2}{2 k}\right)-N (2 N+1) \left(\frac{3}{2}-\frac{2 N+2}{2 k}\right)\nonumber \\
=&\frac{1}{2} \left(\frac{3}{2}-\frac{2 k-2 N}{2 k}\right) (2 k-2 N-2) (2 k-2 N-1)\\
&-\frac{1}{2} \left(\frac{3}{2}-\frac{2 k-2 M-2 N}{2 k}\right) (2 k-2 M-2 N-2) (2 k-2 M-2 N-1) \nonumber
\end{align}
Which, importantly, both eventually expand and reduce to the expression detailed above.

For these cases, we can, as in \cite{Armoni:2015jsa}, also equate the Witten indices of electric and magnetic theories, which count the number of supersymmetric vacua in each theory. They can be obtained combinatorially from the brane picture, counting the number of distinct ways of connecting various fivebranes with threebranes or three-orientifolds: specifically one needs to enumerate the ways of connecting the threebranes to the composite fivebranes, and the number of ways to choose branes from the bigger stack to make the smaller stack. Hence, for the first pair, based on the $USp$ cosets:

\begin{align}
\binom{2k}{2N+2M+1}\binom{2N+2M+1}{2N+1}&=\binom{2k}{2k-2N-1}\binom{2k-2N-1}{2k-2N-2M-1}\\
&=\frac{(2k)!}{(2k-2N-2M-1)!(2N+1)!(2M)!}
\end{align}
\clearpage
And for the second pair, based on the $SO$ cosets:

\begin{align}
\binom{2k}{2N+2M-1}\binom{2N+2M-1}{2N-1}&=\binom{2k}{2k-2N+1}\binom{2k-2N+1}{2k-2N-2M+1}\\
&=\frac{(2k)!}{(2k-2N-2M+1)!(2N-1)!(2M)!}
\end{align}

As was observed in \cite{Armoni:2015jsa}, these formulae provide a further check of the results in the limiting case $M=0$, when the coset is trivial and the entire group becomes gauged. The other limiting case is $2k=2M+2N\pm1$, but this is not allowed by parity.

\subsection{The non-supersymmetric dualities}

While there is no statement akin to the Witten index in these cases, the central charges still do match using the formulae derived above, with the extra shifts in the level due to the imbalance in number of fermions and bosons. The intermediary bosonic and fermionic sector central charges are shown to illustrate that only the end result of the computation matches on either side. Again, there should a priori be no reason why they match, there are no trivial cancellations that happen in the steps of the derivation, only at the very end, with all of the terms present, do the expressions match precisely. Note that for the Type 0'B theories we have taken into account the doubling of the fermion degrees of freedom.

\begin{table}[!h]
\centering
\begin{tabular}{|c|c|c|}
\hline Coset  & \phantom{\LARGE{$\frac{U}{V}$}}\large{$\frac{USp(2N+2M)_{2k}}{USp(2N)_{2k}}$ }&\large{$\frac{USp(2k-2N+2)_{2k}}{USp(2k-2N-2M+2)_{2k}}$}   \\ 
\hline  \multirow{2}{*}{$c_B(R_B)$}& \phantom{\LARGE{$\frac{U}{V}$}} $\frac{M}{2+k}\left(1 + k + M + 2 k M - 2 M^2 \right.$ & $-\frac{M}{2+k}\left(3 + 2 k^2 + 2 M^2 + k (5 - 4 M - 8 N)\right.$ \\
  & $\left.+2 N + 4 k N - 6 M N - 6 N^2 \right)$  & $\left. -10 N + 6 N^2 + M (-5 + 6 N) \right)$ \\
\hline $c_F(R_F)$ & \phantom{\LARGE{$\frac{U}{V}$}} $\frac{1}{2} M (-1 + 2 M + 4 N)$ & $ \frac{1}{2} M (3 + 4 k - 2 M - 4 N)$ \\
\hline \phantom{\LARGE{$\frac{U}{V}$}} $c_{\text{tot}}$  & \multicolumn{2}{c|}{$\frac{M}{4+2k}\left(k (1 + 6 M + 12 N) - 2 (2 M^2 + 6 (-1 + N) N + M (-3 + 6 N)) \right) $}\\
\hline
\end{tabular}
\caption{Matching of the central charges for the first non-supersymmetric pair.}
\end{table}

\begin{table}[!h]
\centering
\begin{tabular}{|c|c|c|}
\hline Coset  & \phantom{\LARGE{$\frac{U}{V}$}}\large{$\frac{SO(2N+2M)_{2k}}{SO(2N)_{2k}}$ }&\large{$\frac{SO(2k-2N-2)_{2k}}{SO(2k-2N-2M-2)_{2k}}$}   \\ 
\hline  \multirow{2}{*}{$c_B(R_B)$}& \phantom{\LARGE{$\frac{U}{V}$}} $-\frac{M}{-2+k}\left(-1 + k + M - 2 k M + 2 M^2 \right.$ & $-\frac{M}{-2+k}\left(3 + 2 k^2 + 2 M^2 + 10 N + 6 N^2\right.$ \\
  & $\left.+2 N - 4 k N + 6 M N + 6 N^2 \right)$  & $\left. +M (5 + 6 N) - k (5 + 4 M + 8 N) \right)$ \\
\hline $c_F(R_F)$ & \phantom{\LARGE{$\frac{U}{V}$}} $\frac{1}{2} M (1 + 2 M + 4 N)$ & $-\frac{1}{2} M (-4 k+2 M+4 N+3)$ \\
\hline \phantom{\LARGE{$\frac{U}{V}$}} $c_{\text{tot}}$  & \multicolumn{2}{c|}{$-\frac{M}{-4+2k}\left(k (1 - 6 M - 12 N) + 2 (2 M^2 + 6 N (1 + N) + M (3 + 6 N)) \right) $}\\
\hline
\end{tabular}
\caption{Matching of the central charges for the second non-supersymmetric pair.}
\end{table}

\begin{table}[!h]
\centering
\begin{tabular}{|c|c|c|}
\hline
Coset & \phantom{\LARGE{$\frac{U}{V}$}}\large{$\frac{U(2N+2M)_{2k}}{U(2N)_{2k}}$} & \large{$\frac{U(2k-2N+2)_{2k}}{U(2k-2N-2M+2)_{2k}}$} \\
\hline  \multirow{2}{*}{$c_B(R_B)$}& \phantom{\LARGE{$\frac{U}{V}$}} $\frac{M}{1+k}\left(1 - 4 M^2 - 12 N^2 \right.$ & $-\frac{M}{1+k}\left(3 + 4 k^2 + 4 M^2 - 16 N + 12 N^2\right.$ \\
  & $\left. +4 M (1 + k - 3 N) + 8 (1 + k) N \right)$  & $\left.-8 k (-1 + M + 2 N) + 4 M (-2 + 3 N) \right)$ \\
\hline $c_F(R_F)$ & \phantom{\LARGE{$\frac{U}{V}$}} $ M (-1 + 2 M + 4 N)$ & $ M (3 + 4 k - 2 M - 4 N)$ \\
\hline
$c$ &  \multicolumn{2}{c|}{\phantom{\LARGE{$\frac{U}{V}$}}$-\frac{M}{1+k}\left( k (1 - 6 M - 12 N) + 4 M^2+ 12 (-1 + N) N + 2M (-3 + 6 N) \right)$}  \\
\hline
\end{tabular}
\caption{Matching for the Type 0'B with antisymmetric fermions}
\end{table}

\begin{table}[!h]
\centering
\begin{tabular}{|c|c|c|}
\hline
Coset &\phantom{\LARGE{$\frac{U}{V}$}}\large{$\frac{U(2N+2M)_{2k}}{U(2N)_{2k}}$} & \large{$\frac{U(2k-2N-2)_{2k}}{U(2k-2N-2M-2)_{2k}}$}\\
\hline  \multirow{2}{*}{$c_B(R_B)$}& \phantom{\LARGE{$\frac{U}{V}$}} $-\frac{M}{-1+k}\left(-1 + 4 M^2 + 12 N^2 \right.$ & $-\frac{M}{-1+k}\left(3 + 4 k^2 + 4 M^2 + 16 N + 12 N^2\right.$ \\
  & $\left. -4 M (-1 + k - 3 N) - 8 (-1 + k) N \right)$  & $\left. -8 k (1 + M + 2 N) + 4 M (2 + 3 N) \right)$ \\
\hline $c_F(R_F)$ & \phantom{\LARGE{$\frac{U}{V}$}} $ M (1 + 2 M + 4 N)$ & $  M (-3 + 4 k - 2 M - 4 N)$ \\
\hline
$c$& \multicolumn{2}{c|}{\phantom{\LARGE{$\frac{U}{V}$}}\phantom{\LARGE{$\frac{U}{V}$}} $ \frac{M}{-1+k}\left( k (1 + 6 M + 12 N) - 4 M^2 -12 N (1 + N) - 2M (3 + 6 N))\right)$}\\
\hline

\end{tabular}
\caption{Matching for the Type 0'B with symmetric fermions}
\end{table} 
\clearpage
\subsection{Stability of the non-supersymmetric set-ups with an excess of fermions}\label{stab}

There is one caveat to the dualities we present. Imposing an imbalance between fermions and bosons of one multiplet will stop certain loop effects from cancelling, naturally. In the cases where fermions are in a symmetric two-index representation, there is an excess of them, typically driving towards negative values the one-loop corrections to the action. This can lead to difficulties: for a charged fundamental scalar, this net excess of fermions would drive its one-loop mass-squared to negative values, i.e. it is expected to grow a vacuum expectation value, breaking the gauge symmetry. In the string picture, this is seen through a repulsive potential between the branes, which then move along a common direction of its bounding fivebranes, to reach a new equilibrium \cite{Sugimoto:2012rt}.  

We must then wonder if, in those theories with a two-index scalar and an excess of fermions, if the former is stabilised or destabilised. In the $(1,1)$ or $(2,0)$ brane configurations this scalar is expected to parametrise the $D3$ brane positions, and is therefore indicative of the potential they observe. We wish to argue that the field theory is still consistent despite this. This is not enough to ensure that the brane picture itself is well-posed, and in general one would need to do a more complex computation to show that.

The states of the boundary degrees of freedom, we argue, have zero measure within the set of all states in three dimensions, so if any instability is detected it should be in the higher-energy regimes of the fully three dimensional theory. In all generality, we should consider coupling a Yang-Mills term to the Chern-Simons fields, since YM-CS theories have a pure CS theory as their low-energy effective action, this will be the most general gauge invariant renormalisable action. Thus we consider the following

\begin{align}
S&_{\text{YMCS+mat.}}=\int_M d^3x \Tr \left( \frac{\kappa}{2}\left( \epsilon^{\mu\nu\rho}\left( A_\mu\partial_\nu A_\rho+\frac{2}{3}A_\mu A_\nu A_\rho\right) -\bar{\lambda}\lambda+2D\sigma\right)\right.\nonumber \\
&\left.+\frac{1}{2g^2}\left( -\frac{1}{2}F^{\mu\nu}F_{\mu\nu}+i\bar{\lambda}\slashed{\mathcal{D}}\lambda+\left( \slashed{\mathcal{D}}\sigma\right) \left( \slashed{\mathcal{D}} \sigma\right) +D^2+i\bar{\lambda}\left[\sigma,\lambda \right]+\left[\sigma,\sigma \right]\left[ \sigma,\sigma\right]   \right)\right)
\end{align}
Where the fields $(A,\lambda,\sigma,D)$ would form a vector multiplet were it not for representation differences, and $\slashed{\mathcal{D}}$ is the covariant derivative. Because all fields are in a double-index representation, they act on each other by the adjoint action, i.e. $\slashed{\mathcal{D}}$ contains a commutator, as does the Yukawa term.  

The presence of a boundary in this computation will not affect the result significantly. In general a boundary, along with appropriate conditions taken there, will affect along which basis of modes we expand in, which imposes further properties of amplitudes, which will not be relevant here.

All fields in the above action are massive, the fermion by an explicit mass term, the scalar by the $D$-term potential, and the gauge field in a non-trivial way that is specific to YM-CS theories. The Chern-Simons term changes the Lorentz structure of the gauge propagator, but also generates a pole at $p^2=g^4\kappa^2$ \cite{Dunne:1998qy}. This peculiarity of such theories enables us to use dimensional regularisation, instead of a UV cutoff, to determine the form of the corrections at hand.

We then wish to compute the mass correction to the field $\sigma$ due to loop effects from $A$ and $\lambda$. As these fields are in mixed representations, which lead to non-trivial commutators between them, it will be beneficial to employ 't Hooft's double-index rules to draw the relevant diagrams. The procedure is to "fatten" each propagator of the fields above, to two lines instead of one, each bearing a colour index rather than a Lie algebra index. Colour indices, like momentum, are expected to "flow" consistently across the diagram, leading to the correct index structure at the end, similarly, a colour loop sums over all indices, resulting in extra factors of $N$. To enforce the definite symmetry of the fields in question, one must add or subtract the diagrams obtained from the original one by twisting one or more of the double-lines. This procedure is, of course, completely equivalent to keeping the relevant fields unexpanded along Lie algebra generators (i.e. writing them as matrices) and deriving the relevant integrals by direct computation, it serves as a helpful book-keeping tool for the various contractions of the colour indices and the symmetries between them that would naturally occur in the process, in a completely analogous fashion to the way Feynman diagrams usually keep track of momentum flow.

Since the only difference between these theories and a supersymmetric one is representation theory, one expects that the scalar part of the corrections (i.e. just the momentum integrals after Lorentz algebraic manipulations) are identical, and that the only difference will come from the aforementioned colour structure. For this purpose, we will perform the fermionic contribution only, since from there one need only modify the computation a small amount in the non-planar diagrams sector to arrive at the bosonic result.

\begin{figure}[h]
\centering
\includegraphics[width=0.98\textwidth]{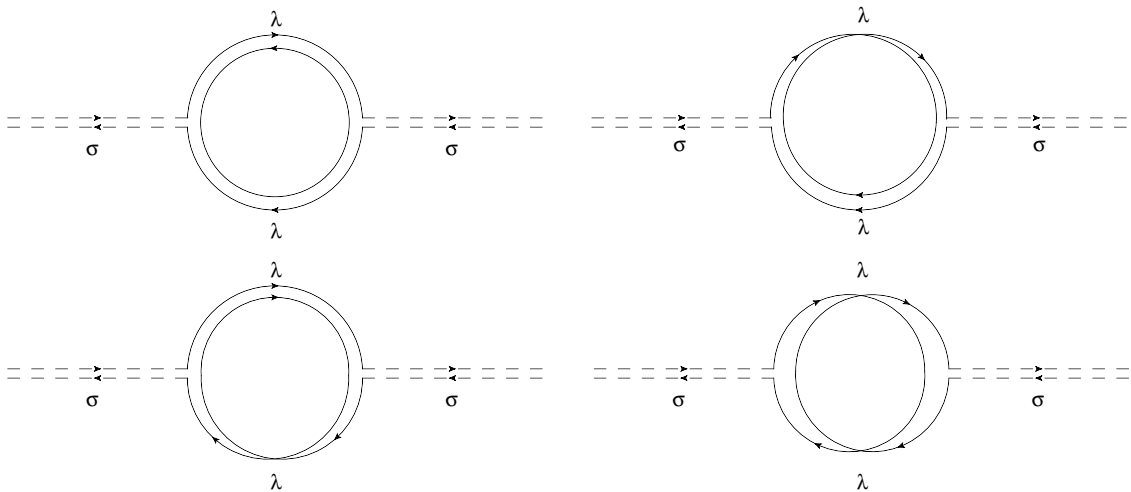}
\caption{Two-index fermionic graphs in the 't Hooft prescription for the gauge scalar mass.}
\end{figure}

We can now read off results from the graphs. For the non-supersymmmetric Type IIB theories one finds that the one-loop quadratic term for $\sigma$ due to the bosonic (resp. fermionic) sector is of the form

\begin{equation}
\pm \sigma_{ij}\left(2N\delta^{il}\delta^{jk} \mp 2\delta^{il}\delta^{jk} +\delta^{ij}\delta^{kl} \right) \sigma_{kl}\times \mathcal{I}(0)
\end{equation}

Where $\mathcal{I}$ is the scalar momentum integral which we will compute shortly. Note the final, colour-diagonal term: it is disconnected, thus discarded in the 1-loop effective action prescription.

\begin{equation}
M^2_\sigma=-4\mathcal{I}(0)=-4g^2\int \frac{d^3p}{(2\pi)^3}\frac{1}{p^2+g^4\kappa^2}=-\frac{4g^2}{4\pi}\sqrt{g^4\kappa^2}=-\frac{g^4|\kappa|}{\pi}
\end{equation}

This quantity, though negative, is finite and $\frac{m^2_{1l}}{m^2_{tree}}=\frac{1}{\pi |\kappa|}$. Now, in a non-Abelian gauge theory, $4\pi \kappa=2k \in2\mathbb{Z}$ \cite{Dunne:1998qy}, which in the above means that for $k>2$ the theory is stable. This is most always the case however: so that our dualities make sense, that is, for the magnetic theory to be well-defined, we must have that $2k\geq2N+2M+2$, so that at most one of $N,M$ can be zero. If both are zero then there is no gauge theory at all, so the above computations make little sense. Therefore, for all values of $k$ of interest, the one-loop tachyonic mass is smaller than the tree level one. Further loop orders will have more factors of $g^2$ before integration, but since $g^2$ has dimension one, by dimensional analysis the integrals will necessarily involve lower and lower (eventually negative) powers of the gauge mass, the upshot of which is that higher loops scale are suppressed by larger powers of $\kappa$ when compared to the tree-level mass.

For the Type 0'B theories: the gauge bosons are in a unitary representation, so yield only a planar diagram, cancelling the $N$-dependent term above. However, since these theories have Dirac fermions instead of Majorana, the end result is that the same mass is generated.

We must also ensure that the $\mathcal{N}=1$ theory is free from such effects. Indeed, we have omitted to mention that the construction detailed previously should have an extra scalar multiplet, given that the bounding fivebranes have a common direction $x^3$. However, it was argued in \cite{Armoni:2015jsa} that it decouples from the parts of the theory that interest us, as the non-abelian components of the theory gain a mass quantum mechanically, a highly non-trivial phenomenon described in \cite{Armoni:2006ee}. It is therefore worth investigating whether the repulsive force detailed here overcomes this attraction or not. The potential found in \cite{Armoni:2006ee} eventually subsumes to giving a mass term of order $g^4 k^2 N^3$ and hence the repulsion computed above is subleading. The extra scalar is, therefore, massive as before and does not destabilise the theory.

Finally, what of the scalar particles in the boundary degrees of freedom? A direct computation of the potential for the relevant scalar is difficult to perform in this case, and we will not do so. We will simply posit that, being fields in a lower-dimensional field theory, the space of such field configurations is necessarily much smaller than the space of field configurations of the three dimensional field (in a measure theoretic sense), and as such we expect the contribution of the former to be naively sub-leading to the effect described here, but, there could be effects we have not accounted for. For these reasons we believe that the dualities between theories with a net excess of fermions make sense, at least when viewed field-theoretically.
\clearpage

\section{Conclusion and outlook}

In the spirit of previous works, we have attempted to generalise Seiberg duality of Wess-Zumino-Witten theories to a wide range of new cases. Starting from Type IIB Hanany-Witten-like setups, we introduced the use of several types of orientifold planes to generate different kinds of theories, particularly in conjunction with anti-branes to break all of supersymmetry, or moving to Type 0'B strings. We have argued that the standard arguments justifying Seiberg-like dualities are still valid, namely that the fivebrane swapping mechanism is still consistent, and that it will lead to new dualities. Thus we support their form from results in string theory. We then provided forms of evidence to justify these dualities. Importantly, the matching of the central charges on either side of the duality provided a strong check of the result: the charges in question are complicated functions of the ranks and levels, only in the very final result is there any semblance of correspondence between the two theories. We have also underlined the importance of the form of the central charges themselves: for the non-supersymmetric theories, these charges bear corrections that are idiosyncratic, specifically characteristic of the altered fermionic representations at hand.

So far, only the above tools have been used to characterise this type of duality, it would be interesting, as a further extension of these results, to test the correspondence for other correlations of operators of the theory, or with further deformations and enhancements (for instance, by adding flavour symmetry and fundamentally charged particles to the theory at hand). These notions and methods could hopefully be used in other settings, also: indeed they are closely related to the ABJM model \cite{Faizal:2011cd} and M2-M5 junctions in general \cite{Berman:2009kj}. It would be interesting to lift these results to analogous situations in a more general setting like the one M-theory provides, as the composite fivebrane we use is more naturally defined in M-theory, it is possible that deeper insight into the phenomena described above could be reached with such a lift. Another sector of interest are domain walls of four-dimensional theories, as they have been argued to bear specific Chern-Simons theories on them \cite{Armoni:2009vv}, most likely there is more to be said about those scenarios. Seiberg duality was originally conceived in four dimensions and more is known about it in that context, and the inclusion of domain walls enables us to probe interesting non-perturbative behaviours of field theory. It could even be brought further by considering junctions of such domain walls, which then behave like the defects we have discussed and possess WZW theories living on them \cite{Gaiotto:2013gwa}.

\textbf{Acknowledgements:} We would like to warmly thank Professors Adi Armoni and Vasilis Niarchos for many useful discussions.

\clearpage
\bibliographystyle{JHEP}
\bibliography{wzw}

\end{document}